# Preparation of $LaO_{0.9}F_{0.1}FeAs$ wires by the powder-in-tube method

Zhaoshun Gao, Lei Wang, Yanpeng Qi, Dongliang Wang, Xianping Zhang, Yanwei Ma*

Key Laboratory of Applied Superconductivity, Institute of Electrical Engineering,

Chinese Academy of Sciences, P. O. Box 2703, Beijing 100190, China

**Abstract:**

We report that Fe sheathed $LaO_{0.9}F_{0.1}FeAs$ wires with Ti as a buffer were fabricated by the powder-in-tube (PIT) method for the first time. Comparing to the common two-step vacuum quartz tube synthesis method, the PIT method is more convenient and safe for synthesizing the novel iron-based layered superconductors. Structural analysis by mean of x-ray diffraction shows that $LaO_{0.9}F_{0.1}FeAs$ is the main phase in wires produced by this synthesis method. The transition temperature of the $LaO_{0.9}F_{0.1}FeAs$ wire is around 25 K. The micrograph shows a homogeneous microstructure with a grain size of 1~3 micrometers. The results suggest that the PIT process is promising in preparing iron-based layered superconductors.

* Author to whom correspondence should be addressed; E-mail: ywma@mail.iee.ac.cn

The recently discovered quaternary arsenide oxide superconductor La[$O_{1-x}F_x$]FeAs with the superconducting critical transition temperature ($T_c$) of 26 K [1], has triggered great interest in both fundamental studies and practical applications due to its higher $T_c$, layered structure and iron-containing character [2-4]. Immediately after this announcement, $T_c$ has been enhanced up to values above 50 K by the substitution of La by smaller rare earth elements [5-10], these compounds thus constituting a new High-$T_c$ family. The high $T_c$ value and the very high upper critical fields ($H_{c2}$) [11, 12] in iron-based layered superconductor have demonstrated that these materials may be competitive with A15, $MgB_2$ and even high-$T_c$ cuprate conductors. However, many critical issues relevant for practical applications remain to be investigated. In particular, iron-based layered superconductors are mechanically hard and brittle and therefore difficult to draw into the desired wire geometry. For wire fabrication, it is necessary to find a suitable sheath material for iron-based layered superconductor, which does not degrade the superconducting properties. In spite of the very high reaction temperature about 1200$^{o}$C of iron-based superconductors, the sheath materials should not react with raw materials and should have a high melt point. Nb or Ta tube may be good candidates, but constitute an expensive solution. Fe is a very cheap metal with high melt point. If we choose an appropriate buffer, it can act as sheath material for iron-based layered superconductor wires. In this paper we present a remarkably simple method for the synthesis of $LaO_{1-x}F_x$FeAs wires using cheap Fe sheath with Ti buffer. The superconducting properties and microstructural features of $LaO_{1-x}F_x$FeAs wires were also investigated.

The $LaO_{0.9}F_{0.1}$FeAs composite wires were prepared by the in situ powder-in-tube (PIT) method using La, As, $LaF_3$, Fe and $Fe_2O_3$ as starting materials. The raw materials were thoroughly grounded by hand with a mortar and pestle. The mixed powder was filled into a Fe tube of 8 mm outside diameter and 1.5 mm wall thickness. A Ti sheath with the thickness of 0.03 mm was inserted in order to prevent a reaction between the raw materials and the Fe tube. It has to be noted that the grinding and packing processes were carried out in a glove box under high pure Argon. After packing, the tube was rotary swaged and then drawn to wires of 2.0 mm in diameter. The wires were cut into pieces of 4 to 6 cm length and were sealed in a Fe tube. They are then reacted at 1150 $^{o}$C for 40 hours. The high purity argon gas was allowed to flow into the furnace during the heat-treatment process to reduce the oxidation of the samples. It should be noted that this process is also suitable for preparing iron-based layered superconductors bulks after little modification [13].

The phase identification and crystal structure investigation were carried out using x-ray diffraction (XRD). Standard four-probe resistance and ac magnetization measurements were carried out using a physical property measurement system (PPMS). Microstructure was studied using a scanning electron microscopy (SEM) after peeling away the Fe sheath.

The final sizes of wires are shown in Fig. 1(a). The sheath to core ratio of the fabricated wire is around 1.58. Figures 1 (b) and (c) show the SEM images for a typical transverse and a longitudinal cross-section after heat treatment. It clearly shows that the $LaO_{0.9}F_{0.1}FeAs$ core presents a homogeneous cross-section. We did not observe a reaction between $LaO_{0.9}F_{0.1}FeAs$ and Fe sheath, thus demonstrating the effectiveness of the Ti buffer sheath.

Figure 2 presents the XRD pattern of $LaO_{0.9}F_{0.1}FeAs$ wire after peeling off the Fe sheath. As can be seen, the sample consists mainly of $LaO_{0.9}F_{0.1}FeAs$, but some impurity phases are also detected. These impurity phases could be caused by the inadequate sintering time and temperature [2, 8].

Figure 3 shows the temperature dependences of resistance for $LaO_{0.9}F_{0.1}FeAs$ wire with the measuring current of 1mA. From this figure we can observe a sharp transition with the onset temperature 24.6 K and zero resistance at 20.5 K, indicating a good quality of our samples. According to Fig.3, a residual resistance is observed under $T_c$, which may be due to the reaction layer between the Ti and core materials. The inset shows the temperature dependence of AC magnetization with the measuring frequency of 333 Hz and the amplitude of 0.1 Oe. The onset critical temperature by magnetic measurement is about 21.8 K, which corresponds to the middle transition point of resistance.

Figure 4 shows the magnetic field dependence of the critical current density $J_c$

derived from the hysteresis loop width on the basis of the extended Bean model using the full sample dimensions. The inset of Fig. 4 shows the *M-H* loop at 5 *K* with the ferromagnetic background, which is probably due to unreacted Fe or $Fe_2O_3$ impurity phase, as observed by Flükiger et al. [12]. From Fig. 4 we can see that the $J_c$ is low and very sensitive to the external magnetic field. The low $J_c$-$H$ properties imply either very weak pinning or an imperfectly connected superconducting state [14]. The work for improving the $J_c$-$H$ properties of $LaO_{1-x}F_xFeAs$ wires is under progress.

In order to investigate the microstructure and determine the chemical composition of our samples, the SEM and energy dispersive x-ray (EDX) microanalysis are employed. Figure 5(a) and (b) show the typical SEM images of the fractured core layers for $LaO_{0.9}F_{0.1}FeAs$ wires. It can be seen that the $LaO_{0.9}F_{0.1}FeAs$ has an apparently homogeneous structure with some voids. These voids might be a result of Arsenic gas release from the two ends of wires at high temperature. Quantitative analysis by EDX confirms that the mole ratio among the La, Fe, As is around 1:1:0.75. From the higher magnification image we can find the layered structure feature with the grain size of 1~3 micrometers. This result is similar to samples synthesized by two-step method [15].

We have prepared $LaO_{0.9}F_{0.1}FeAs$ composite wires with a Ti buffer by the remarkably simple PIT process. Microstructure analysis indicates that there is no reaction between $LaO_{0.9}F_{0.1}FeAs$ and Fe sheath. Temperature dependence of the magnetization shows that the onset transition temperature is 24.6 K. Our preliminary results suggest that this process is suitable for producing iron-based layered superconductors.

The authors thank Profs. Zizhao Gan, Huhai Wen, Liye Xiao and Liangzhen Lin for their help and useful discussion. This work is partially supported by the Beijing Municipal Science and Technology Commission under Grant No. Z07000300700703, National ‘973’ Program (Grant No. 2006CB601004) and National ‘863’ Project (Grant No. 2006AA03Z203).

## References

[1] Kamihara Y., Watanabe T., Hirano M. and Hosono H. *J. Am. Chem. Soc.* **130,** 3296 (2008).

[2] Wen H. H., Mu G., Fang L., Yang H., and Zhu X. Y., Europhys. Lett. **82**, 17009 (2008).

[3] Chen G. F., Li Z., Li G.,Zhou J., Wu D., Dong J., Hu W. Z., Zheng P., Chen Z. J. Luo J. L and Wang, N. L., *Condmat:arXiv*, 0803- 0128 (2008).

[4] Lorenz B., Sasmal K., Chaudhury R. P., Chen X. H., Liu R. H., Wu T. and Chu C. W., *Cond-mat:arXiv*, 0804.1582 (2008).

[5] Chen X. H., Wu T., Wu G., Liu R. H., Chen H. and Fang D. F., *Condmat:arXiv*, 0803-3603 (2008).

[6] Chen G. F., Li Z., Wu D., Li G., Hu W. Z., Dong J., Zheng P., Luo J. L. and Wang, N. L., *Condmat:arXiv*, 0803- 3790 (2008).

[7] Ren Z. A., Yang J., Lu W., Yi W., Che G. C., Dong X. L., Sun L. L. and Zhao Z. X., *Condmat: arXiv*, 0803.4283 (2008).

[8] Ren Z. A., Yang J., Lu W., Yi W., Shen X. L., Li Z. C., Che G. C., Dong X. L., Sun L. L., Zhou F. and Zhao Z. X., *Cond-mat:arXiv*, 0803.4234 (2008).

[9] Ren Z. A., Yang J., Lu W., Yi W., Shen X. L., Li Z. C., Che G. C., Dong X. L., Sun L. L., Zhou F. and Zhao Z. X., *Cond-mat:arXiv*, 0804.2053 (2008).

[10] Cheng P., Fang L., Yang H., Zhu X. Y., Mu G., Luo H. Q, Wang Z. S and Wen H. H. *Cond-mat:arXiv*, 0804.0835 (2008).

[11] Hunte F., Jaroszynski J., Gurevich A., Larbalestier D. C., Jin R., Sefat A.S., McGuire M.A., Sales B.C., Christen D.K., Mandrus D., *Cond-mat:arXiv*, 0804.0485 (2008).

[12] Senatore C., Flükiger R., Cantoni M., Wu G., Liu R. H. and Chen X. H. *Cond-mat:arXiv*, 0805.2389 (2008).

[13] Yanwei Ma, Zhaoshun Gao, Lei Wang, Yanpeng Qi, Dongliang Wang, Xianping Zhang, *Cond-mat:arXiv*, 0806.2839 (2008).

[14] Yamamoto A., Jiang J., Tarantini C., Craig N., Polyanskii A. A., Kametani F., Hunte F., Jaroszynski J., Hellstrom E. E., Larbalestier D. C., Jin R., Sefat A. S., McGuire M. A., Sales B. C., Christen D. K., and Mandrus D. *Appl.Phys.Lett.* **92** 252501 (2008).

[15] Lu W., Dong X. L., Ren Z. A., Che G. C. and Z.X. Zhao, *Cond-mat:arXiv*, 0803.4266 (2008).

# Captions

Figure 1  Photograph of the final wires (a). SEM images for a typical transverse (b) and a longitudinal (c) cross-section after heat treatment.

Figure 2  XRD patterns of $LaO_{0.9}F_{0.1}FeAs$ wire after peeling away the Fe sheath. The impurity phases of LaAs and LaOF are marked by * and #, respectively.

Figure 3  Temperature dependences of resistance for $LaO_{0.9}F_{0.1}FeAs$ wire with Fe sheath. The inset shows the temperature dependence of the real part of the AC magnetization.

Figure 4  Magnetic field dependence of $J_c$ at 5 K for $LaO_{0.9}F_{0.1}FeAs$ wires. The inset shows the *M-H* loop at 5 K with the ferromagnetic background.

Figure 5  (a) Low magnification and (b) high magnification SEM micrograph for $LaO_{0.9}F_{0.1}FeAs$ samples. (c) EDS element analysis of $LaO_{0.9}F_{0.1}FeAs$, the rectangle marks the position where we took the EDX spectrum.

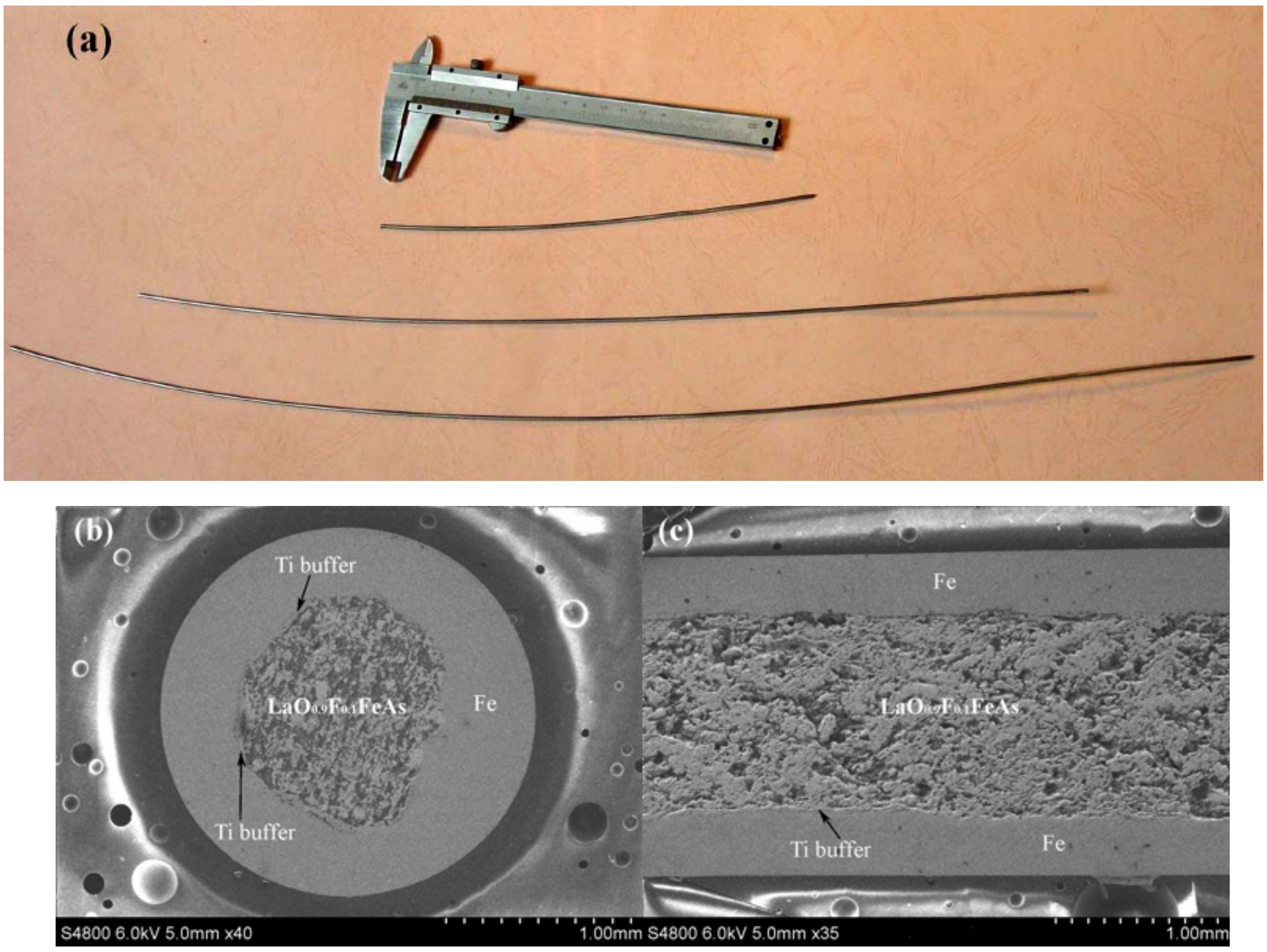

Fig.1 Gao et al.

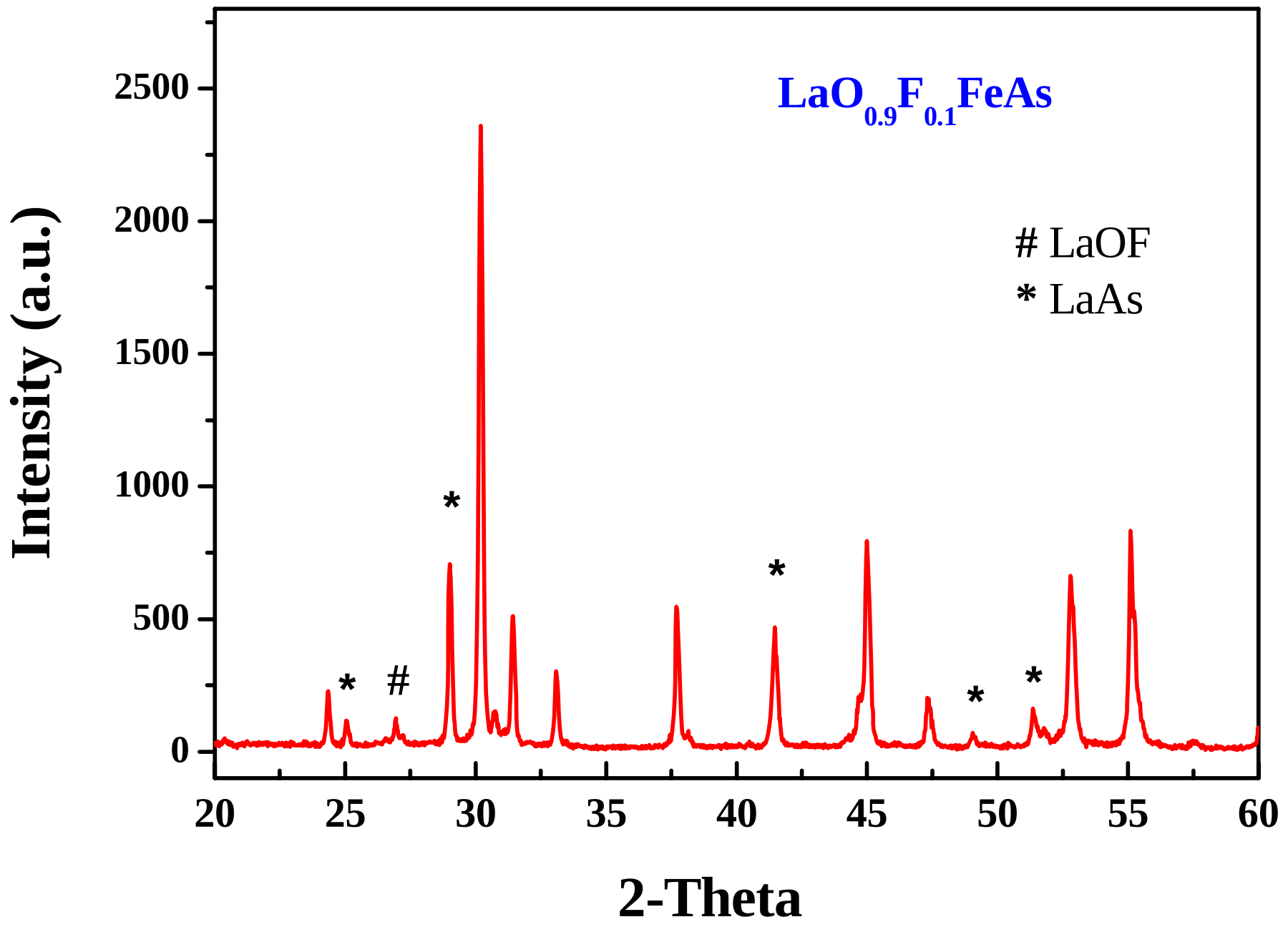

Fig.2 Gao et al.

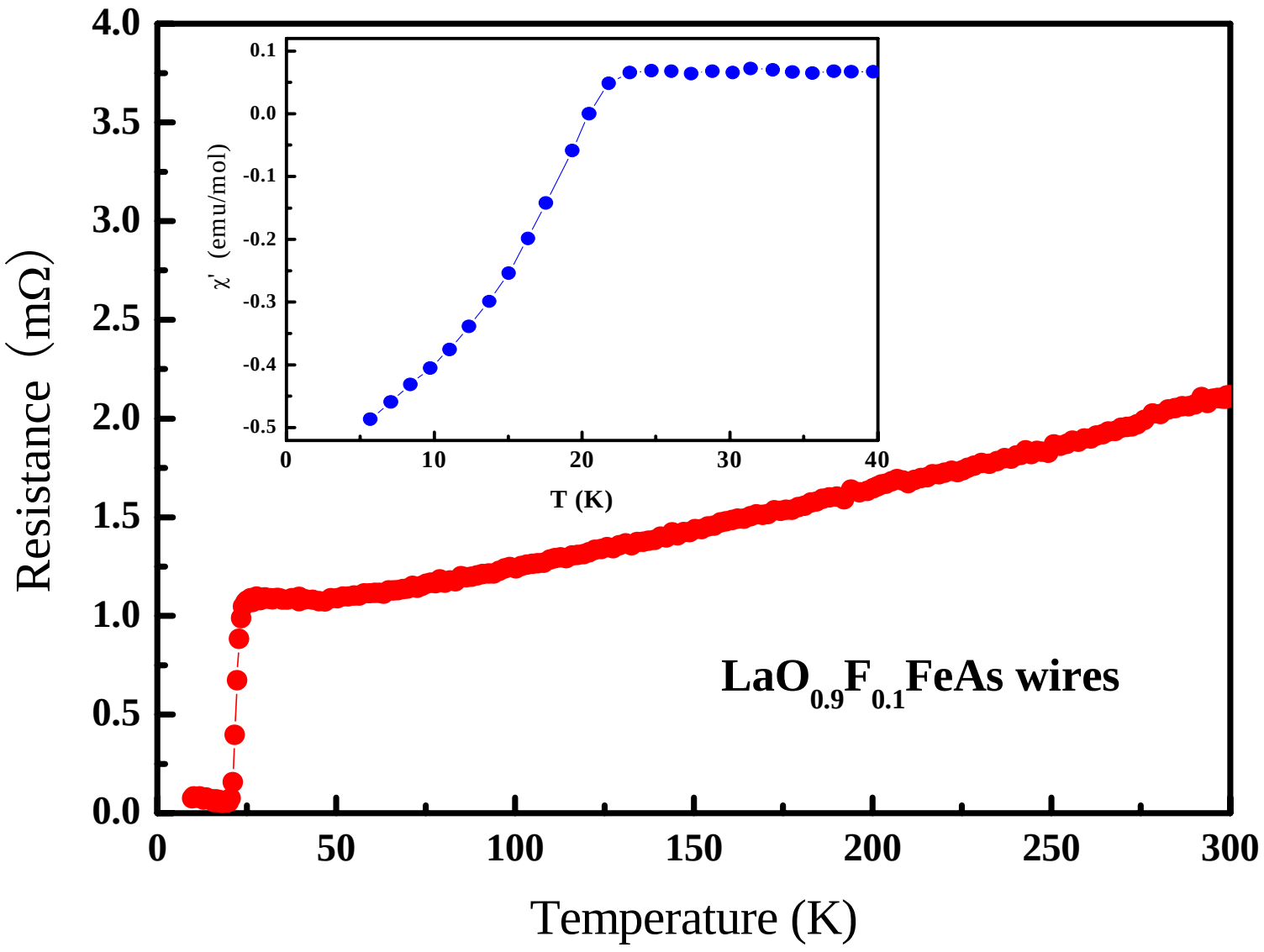

Fig.3 Gao et al.

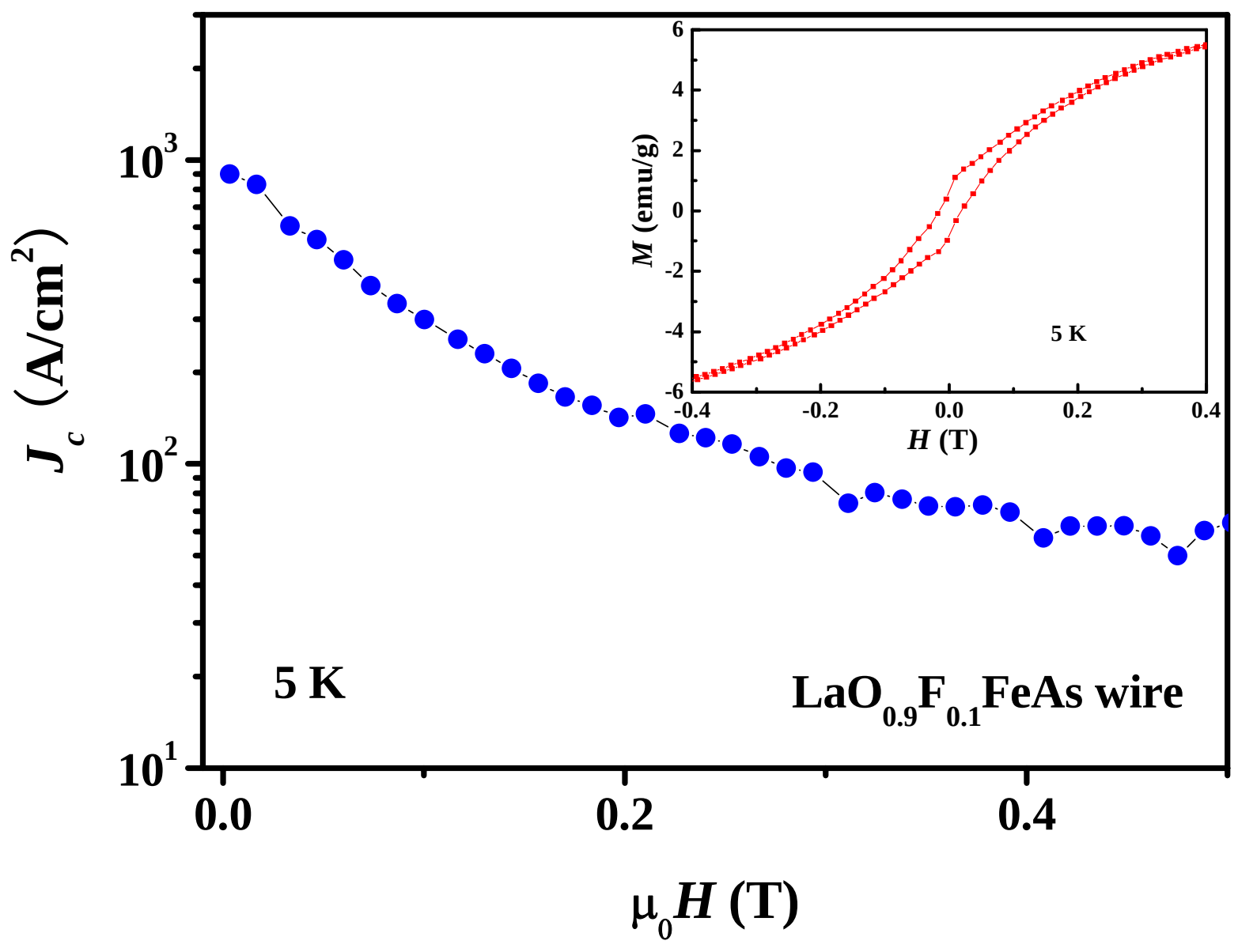

Fig.4 Gao et al.

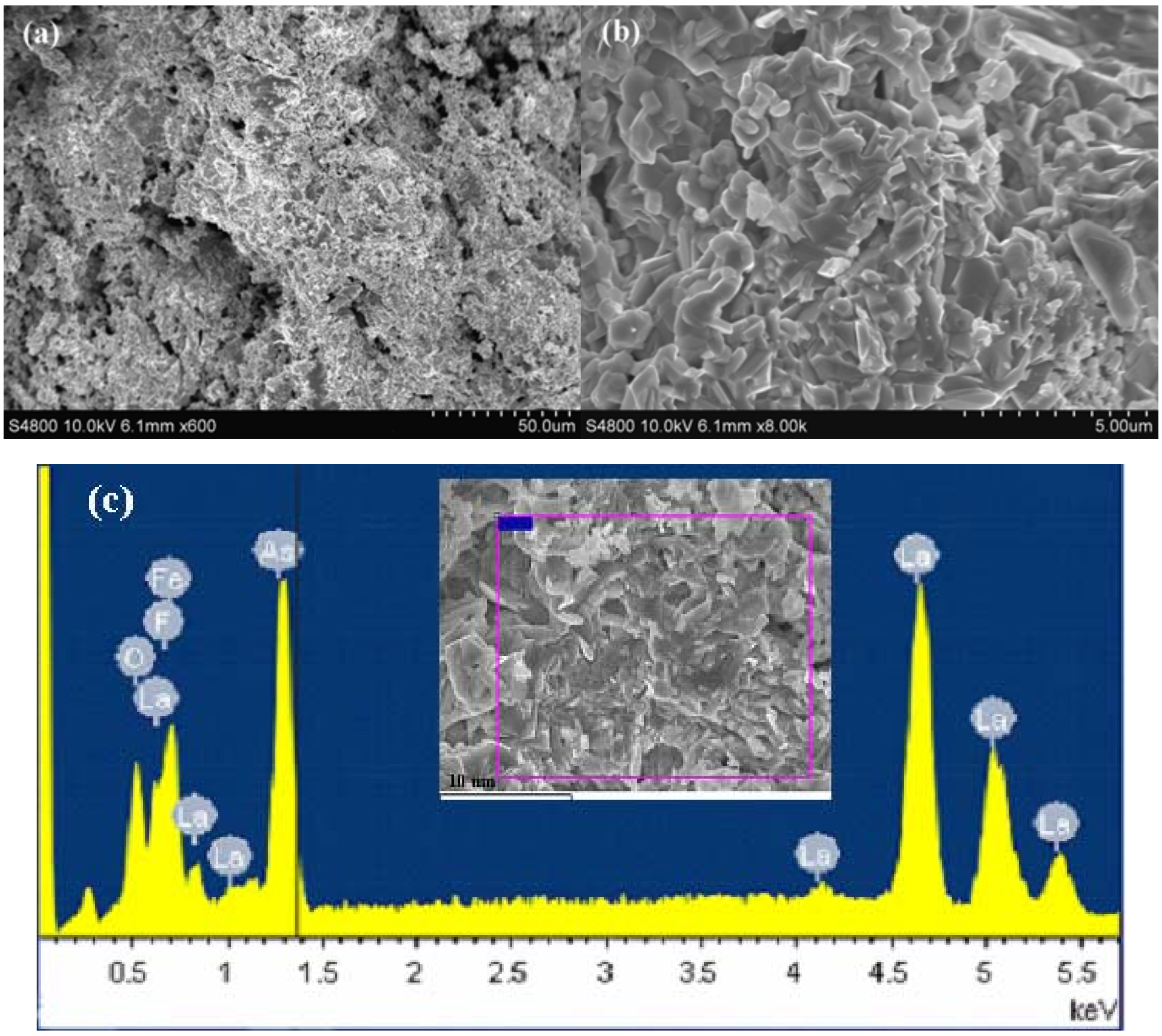

Fig.5 Gao et al.